\documentstyle[manuscript,prl,aps]{revtex}
\begin{document}
\title{Equivalence of the Calogero-Sutherland Model to Free Harmonic 
Oscillators}
\author{N. Gurappa and Prasanta. K. Panigrahi}
\address{School of Physics \\
University of Hyderabad,\\
Hyderabad, Andhra Pradesh,\\
500 046 INDIA}
\maketitle

\begin{abstract} 
A similarity transformation is constructed through which a system of
particles interacting with inverse-square two-body and harmonic potentials
in one dimension, can be mapped identically, to a set of free harmonic
oscillators. This equivalence provides a straightforward method to find
the complete set of eigenfunctions, the exact constants of motion and a
{\it linear} $W_{1+\infty}$ algebra associated with this model. It is also
demonstrated that a large class of models with long-range interactions,
both in one and higher dimensions can be made equivalent to decoupled
oscillators.
\end{abstract}
\draft
\newpage

In recent times, the one-dimensional system of identical particles having
pair-wise inverse-square and harmonic interactions [1], known in the
literature as the Calogero-Sutherland (CS) model, has generated wide
interest. This exactly solvable model and its generalizations to the periodic
case [2] and the spin systems [3], have been found relevant for the
description of various physical phenomena such as the universal conductance
fluctuations in mesoscopic systems [4], quantum Hall effect [5], wave
propagation in stratified fields [6], random matrix theory [2,7], fractional
statistics [8], two-dimensional gravity [9] and gauge theories [10].

Since its inception, this remarkable many-body system, with long-range
interactions  has been studied quite extensively in the literature for a
better understanding of the origin of šsolvability and the underlying
symmetries [11-13]. The energy eigenvalues and the level degeneracies of the
CS model match identically with those of harmonic oscillators, apart from a
coupling dependent shift of the ground-state energy. This structure of the
spectrum has led Calogero to suggest the interesting possibility of a map
between the CS model and decoupled oscillators [14].

In this paper, we provide a similarity transformation which realizes this
correspondence by exactly mapping the CS system of interacting particles,
to a set of free harmonic oscillators. This equivalence provides an
elegant and straightforward method to construct the complete eigenstates of the
CS model [15], including the degenerate ones, starting from the
symmetrized form of the eigenstates of the harmonic oscillators and
reveals the existence of a {\it linear} $W_{1+\infty}$ algebra as the
infinite dimensional symmetry associated with the CS system. It also
allows one to determine the $N$ linearly independent, mutually commuting
constants of motion and demonstrate that, a large class of CS type models
in one and higher dimensions can also be solved in an analogous manner.

The $N$ particle CS Hamiltonian is given by (in the units 
$\hbar=\omega=m=1$)
\begin{equation}
H = - \frac{1}{2} \sum_{i=1}^N \frac{{\partial}^2}{\partial x_i^2}
+ \frac {1}{2} \sum_{i=1}^N x_i^2 + \frac {1}{2} g^2 \sum_{{i,j=1}\atop
{i\ne j}}^N \frac {1}{(x_i - x_j)^2}\qquad,
\end{equation}
here, $g^2 > - \frac{1}{4}$ is the coupling constant. We work in a sector
of the configuration space corresponding to a definite ordering of the
particle coordinates: $x_1 \le x_2 \le \cdots \le x_N$. This is possible
in one dimension since the particles can not overtake each other in the
presence of the repulsive interaction. The resulting wave function is then
analytically continued to other sectors of the configuration space, in
order to obtain the actual eigenfunction.

The correlated ground-state of $H$ is known to be of the form $\psi_0 = Z G$,
where $Z \equiv \prod_{i<j}^N [|x_i - x_j|^\alpha (x_i - x_j)^\delta]$ and $G
\equiv \exp\{-\frac{1}{2}  \sum_i x_i^2\}$. The following similarity
transformation then yields 
\begin{equation}
\tilde H \equiv \psi_0^{-1} H \psi_0 = \sum_i x_i \frac{\partial}
{\partial x_i} + E_0 - \hat A \qquad, 
\end{equation}
where, $E_0 = \frac{1}{2} N + \frac{1}{2} N (N-1) (\alpha + \delta)$ is
the ground-state energy and $\hat A \equiv[\frac{1}{2}
\sum_i\frac{\partial^2} {\partial x_i^2} + (\alpha + \delta) \sum_{i\ne j}
\frac{1}{(x_i - x_j)}
\frac{\partial}{\partial x_i}]$. The eigenfunctions of $\tilde H$ must be
totally symmetric with respect to the exchange of any two particle
coordinates; the bosonic or the fermionic nature of the wave function
being contained in the Jastrow factor $Z$. One can also show that,
\begin{equation}
\left[\tilde H , \exp\{- \hat A/2\}\right] = 
\left[\sum_i x_i \frac{\partial}{\partial x_i} , \exp\{- \hat A/2\}\right] 
= \hat A \exp\{- \hat A/2\} \qquad.
\end{equation}
Making use of (3) in (2), it can be easily verified that, the operator
$\hat T \equiv Z G \exp\{- \hat A/2\}$ diagonalizes the original
Hamiltonian $H$:
\begin{equation}
{\hat T^{-1}} H \hat T = \sum_i x_i \frac{\partial}{\partial x_i} + E_0 
\qquad.
\end{equation}
The above equation is also obtained by Sogo [16]. The following
similarity transformation on (4) achieves the correspondence of the CS
model with the decoupled oscillators;
\begin{equation}
G \hat E {\hat T}^{-1} H \hat T {\hat E}^{-1} G^{-1} =
-\frac{1}{2} \sum_i \frac{\partial^2}{\partial x_i^2} + \frac{1}{2} \sum_i
x_i^2 + (E_0 - \frac{1}{2} N) \qquad,
\end{equation}
where, $\hat E = \exp\{-\frac{1}{4}\sum_i \frac{\partial^2} {\partial
x_i^2}\}$. As anticipated by Calogero and indicated by the structure of the
eigenspectrum, only the ground-state energy depends on the coupling constant;
the rest of the Hamiltonian describes $N$ free oscillators. Hence,
it follows that, the excited energy levels and the degeneracy
structure of both the systems are identical, a fact known since the original
solution of the interacting model [1].

From (4), one can define the creation and annihilation operators of CS
model as $a_i^+ = \hat T x_i {\hat T}^{-1}$ and $a_i^- = \hat T \partial_i
{\hat T}^{-1}$: $[a_i^-\,\,,\,\,a_j^+] = \delta_{ij}$ and the CS
Hamiltonian becomes
\begin{equation}
H = \sum_i H_i = \frac{1}{2} \sum_i \{a_i^-\,\,,\,\,a_i^+\} 
\qquad,
\end{equation}
where, $H_i \equiv \frac{1}{2} \{a_i^-\,\,,\,\,a_i^+\} + \frac{1}{2}
[\alpha (N - 1)]$ and $[H_i\,\,,\,\,a_i^- (a_i^+)] = - a_i^- (a_i^+)$.

One can also define $<<0|{S}_n(\{a_i^-\}) = <<n|$ and ${S}_n(\{a_i^+\})|0>
= |n>$ as the bra and ket vectors; ${S}_n$ is a symmetric and homogeneous
function of degree $n$ and $<<0|a_i^+ = a_i^- |0> = 0$. Since the
oscillators are decoupled, the inner product between these bra and ket
vectors proves that any ket $|n>$, with a given partition of $n$, is
orthogonal to all the bra vectors, with different $n$ and also to those
with different partitions of the same $n$.

For the construction of the eigenfunctions of CS model, one can also make use
of (4) since, $x_i$ and $\frac{\partial}{\partial x_i}$ serve as the creation
and annihilation operators respectively. The ground-state can be chosen:
$\frac{\partial}{\partial x_i} \phi_0 = 0 \,\,,\,\,\mbox{for}\,\,
i=1,2,\cdots, N$. The excited states are given by the monomials $\prod_l^N
x_i^{n_l}$ taken in a symmetric form, with respect to the exchange of
particle coordinates; here, $n_l = 0, 1, 2,\cdots$ and $E = \sum_l n_l +
E_0$. There exist several related basis sets for these functions, invariant
under the action of the symmetric group $S_N$, viz., Schur functions,
monomial and complete symmetric functions [17]. The eigenstates of CS system
spanning the $n$-th energy level can be written as
\begin{equation}
\Psi_n = \psi_0 [\exp\{-\hat A/2\}\,\,S_n] = \psi_0 \,\, P_n \qquad. 
\end{equation}
Here, $P_n \equiv \exp\{- \hat A/2\} S_n$ are totally symmetric,
inhomogeneous polynomials and $S_n$s are any of the above mentioned
symmetric polynomials. One special choice of $S_n$'s is provided by the
Jack polynomials [17]. This basis set have been recently studied by T.H. Baker
and P.J. Forrester, who also obtain the ground state normalization [15]. 

Although the Hamiltonian consists of decoupled oscillators, it is of deep
interest to note that, the individual particle states created by the
action of the powers of a single $a_i^+$ on the ground state give rise to
functions which contain negative powers of the particle coordinates and
hence is in general, not normalizable. But, the appropriate symmetric
combinations of these non-normalizable functions results in polynomials,
which are normalizable with respect to the ground state wavefunction as
the weight function. This fact reveals the many-particle correlation
inherent in the CS model and allows only the $N$-particle states to belong
to the physical Hilbert space.

The quantum integrablility of the CS model has been proved earlier by
identifying an infinite number of constants of motion [13]. It is however
clear that one should look for {\it only} $N$ linearly independent
conserved quantities, since this is a $N$ particle system in one
dimension. From (6), one can check that $[H\,\,,\,\,H_k] =
[H_i\,\,,\,\,H_j] = 0; i,j,k = 1,2,\cdots, N$. Therefore, the set $\{H_1,
H_2,\cdots, H_N\}$ provides the $N$ such conserved quantities. One can
construct, $N$ linearly independent symmetric conserved quantities from
the elementary symmetric polynomials, since they form a complete set:
\begin{equation}
I_1 = \sum_{1 \le i \le N} H_i \,\,, I_2 = \sum_{1 \le i < j \le N} H_i
H_j \,\,, I_3 = \sum_{1 \le i < j < k \le N} H_i H_j H_k,
\cdots, I_N = \prod_{i=1}^N H_i.
\end{equation}

Here, we would like to point out that, one can also construct the
following type of conserved quantities using $H_i$ and $a_i^+$:
\begin{equation}
I = \sum_i H_i^2 + \alpha \sum_{{i,j} \atop {i \ne j}} \frac {a_i^+ +
a_j^+}{a_i^+ - a_j^+} (H_i - H_j)
\end{equation}
and its eigenfunctions are $ Z^\alpha G (\exp\{- A/2\} J_\lambda)$ [15],
which are also the eigenfunctions of the CS Hamiltonian; here, $J_\lambda$
is a Jack polynomial of degree $\lambda$.

Akin to the free oscillator case, one can define a linear $W_{1+\infty}$
algebra for the CS model, using $a_i^-$ and $a_i^+$. We choose one of the
basis for the generators of the $W_{1+\infty}$ as $L_{m,n} = \sum_{i=1}^N
{(a_i^+)}^{m+1} {(a_i^-)}^{n+1}$, for $m, n \ge - 1$ [18]. These will obey
the linear relation
\begin{equation}
[L_{m,n}\,\,,\,\,L_{r,s}] = \sum_{p=0}^{\mbox{Min}(n,r)} \frac{(n+1)!
(r+1)!} {(n-p)! (r-p)! (p+1)!} L_{m+r-p,n+s-p} - (n \leftrightarrow s, m
\leftrightarrow r) \qquad.
\end{equation}
The highest weight vector obtained from $L_{m,n} \psi_0 = 0$ for $n > m
\ge -1$ is nothing but the CS model ground-state wave function. One can
also choose $W_n^{(s)} \equiv L_{n+s-2,s-2}$ for $s \ge 1$ and $n \ge 1 -
s$; where, $W_n^{(s)}$ is the $n$-th Fourier mode of a spin $s$ field.
This result may find application in fractional quantum Hall effect, since
the Laughlin type wave functions can be related to the CS model [5]. The
previously known basis in the literature gave a non-linear realization of
the above algebra, for which the coupling and $\hbar$ dependent non-linear
terms were not known explicitly [13]. We would like to remark that the
above analysis can also be performed for the translationally invariant
model [1].

In the following, we give an example of an interacting many-body
Hamiltonian which can be solved in a manner analogous to the CS model.

We consider a recently studied two-dimensional model [19]:
\begin{equation}
H = -\frac{1}{2} \sum_i \vec{\nabla}_i^2 + \frac{1}{2} \sum_i \vec{r}_i^2
+ \frac{g_1}{2} \sum_{i,j \atop {i\ne j}} \frac{\vec{r}_i^2}{Y_{ij}^2} +
\frac{g_2}{2} \sum_{i,j,k \atop {i\ne {j\ne k}}} \frac{\vec{r}_j}{Y_{ij}} 
\frac{\vec{r}_k}{Y_{ik}} \qquad.
\end{equation}
Here, $\vec{r}_i^2 = x_i^2 + y_i^2$ and $Y_{ij} = x_i y_j - x_j y_i$. The
operator which brings the above Hamiltonian to the diagonal form {\it
i.e.}, $\hat U^{-1}\,\, H\,\, \hat U = \sum_i x_i \frac{\partial}{\partial
x_i} +
\sum_i y_i \frac{\partial}{\partial y_i} + E_0$, is 
given by
\begin{equation}
\hat U = \prod_{i<j}^N |Y_{ij}|^g\,\,\exp\{-\frac{1}{2}\sum_i \vec{r}_i^2\}\,\,
\exp\{- \hat C/2\}\qquad,
\end{equation}
where, $\hat C \equiv \frac{1}{2}\sum_i( \frac{\partial^2}{\partial x_i^2}
+ \frac{\partial^2}{\partial y_i^2}) - g \sum_{i,j \atop
{i\ne j}} \frac{1}{Y_{ij}} (x_j \frac{\partial}{\partial y_i} - y_j
\frac{\partial}{\partial x_i})$, $E_0 = N + g N (N-1)$, $g_1 = g (g-1)$ and
$g_2 = g^2$. One set of excited states and the corresponding energy
spectrum are given by $\Psi_{n,l=0} = \hat U S_n(r_1^2, r_2^2,\cdots,
r_N^2)$ and $E_{n,l=0} = 2 n + E_0 = 2 \sum_{k=1}^N n_k + E_0$
respectively; here, $l$ refers to the angular momentum quantum number and
the symmetric function $S_n(r_1^2, r_2^2,\cdots, r_N^2)$ 
can be constructed from any of the earlier mentioned basis. Another set of
eigenstates for $l \ne 0$ can be constructed by choosing the symmetric
function to be $S_{n_1,n_2} = {(\sum_i z_i^2)}^{n_1} {(\sum_i {\bar
z}_i^2)}^{n_2}$; here, $z_i \,\,(\bar z_i) = x_i + i y_i \,\,(x_i - i y_i)$
and the corresponding energy eigenvalues are $E_{n_1,n_2} = n_1 + n_2 + E_0$. 

As is clear from the above analyses, the $D$ dimensional $N$ particle
Hamiltonians which can be brought through a suitable transformation to the
generalized form: $\tilde H = \sum_{l=1}^D \sum_{i=1}^N x_i^{(l)}
\frac{\partial}{\partial x_i^{(l)}} + c + \hat F$ can also be mapped  
to $\sum_{l=1}^D \sum_{i=1}^N x_i^{(l)}
\frac{\partial}{\partial x_i^{(l)}} + c$ by $\exp\{- d^{-1} \hat F\}$; where,
the operator $\hat F$ is any homogeneous function of
$\frac{\partial}{\partial x_i^{(l)}}$ and $x_i^{(l)}$ with degree $d$ and
$c$ is a constant. For the normalizability of the wave functions, one
needs to check that the action of $\exp\{- d^{-1} \hat F\}$ on an
appropriate linear combination of the eigenstates of $\sum_{l=1}^D
\sum_{i=1}^N x_i^{(l)}
\frac{\partial}{\partial x_i^{(l)}}$ yields a polynomial solution. These
Hamiltonians fall in one class in the sense that, any member of the class
can be mapped to the other via free harmonic oscillators. This result may
find application in other branches of physics and mathematics.

Since, the mapping between the CS model and harmonic oscillators was
established by a similarity transformation, one can immediately construct
the coherent state for CS model [20] starting from the coherent state of
harmonic oscillators. Akin to the harmonic oscillator coherent states,
these will be extremely useful for finding the classical limit of the
quantum CS model [21]. We also remark that, the presence of the
$W_{1+\infty}$ algebra in the oscillator representation indicates the
possibility of having Kadomtsev-Petviashvili (KP) type non-linear
equations in this model [22]. Work along the above lines as well as the
application of our technique to spin Calogero models are currently under
progress and will be reported elsewhere.

In conclusion, we have shown the equivalence of the CS model to $N$ free
harmonic oscillators; this led to a straightforward construction of the
eigenstates, the $N$ conserved quantities responsible for the quantum
integrability of the system and the generators of the {\it linear}
$W_{1+\infty}$ algebra. CS type many-body Hamiltonian in higher-dimension
was also solved in an analogous manner.

The authors would like to acknowledge useful discussions with Profs. V.
Srinivasan, S. Chaturvedi and Asok Das. N.G thanks E. Harikumar for useful
comments and U.G.C (India) for financial support through the S.R.F scheme.


\begin{references}
\item F. Calogero, J. Math. Phys. {\bf 3}, 419 (1971);
      B. Sutherland, {\it ibid}, {\bf 12}, 246 (1971); {\bf 12}, 251 (1971).
\item B. Sutherland, Phys. Rev. {\bf A 4}, 2019 (1971); {\bf 5}, 1372 (1972);
      L. Lapointe and L. Vinet, Commun. Math. Phys. {\bf 178}, 425 (1996).
\item F.D.M. Haldane, Phys. Rev. Lett. {\bf 60}, 635 (1988); B.S. Shastry, 
      {\it ibid}, 639 (1988).
\item S. Tewari, Phys. Rev. {\bf B 46}, 7782 (1992); N.F. Johnson and 
      M.C. Payne, Phys. Rev. Lett. {\bf 70}, 1513 (1993); {\bf 70}, 3523 
      (1993); M. Caselle, {\it ibid}. {\bf 74}, 2776 (1995).
\item N. Kawakami, Phys. Rev. Lett. {\bf 71}, 275 (1993); H. Azuma and S. Iso,
      Phys. Lett. {\bf B 331}, 107 (1994);  P.K. Panigrahi and M. Sivakumar,
      Phys. Rev. {\bf B 52}, 13742 (15) (1995).
\item H.H. Chen, Y.C. Lee and N.R. Pereira, Phys. Fluids {\bf 22}, 187
      (1979).
\item B.D. Simons, P.A. Lee and B.L. Altshuler, Phys. Rev. Lett. 
      {\bf 70}, 4122 (1993); {\bf 72}, 64 (1994).
\item J.M. Leinaas and J. Myrheim, Phys. Rev. {\bf B 37}, 9286 (1988); 
      F.D.M. Haldane, Phys. Rev. Lett. {\bf 67}, 937 (1988); 
      A.P. Polychronakos, Nucl. Phys. {\bf B 324}, 597 (1989);
      M.V.N. Murthy and R. Shankar, Phys. Rev. Lett. {\bf 73}, 3331 (1994).
\item I. Andric, A. Jevicki and H. Levine, Nucl. Phys. {\bf B 312}, 307 (1983);
      A. Jevicki, {\it ibid}, {\bf 376}, 75 (1992).
\item J.A. Minahan and A.P. Polychronakos, Phys. Lett. {\bf B 312}, 155 (1993);
      {\bf 336}, 288 (1994). 
\item For a review see, A.M. Perelomov, Theor. Math. Phys. {\bf 6}, 263
      (1971); M.A. Olshanetsky and A.M. Perelomov, Phys. Rep. {\bf 94}, 6
      (1983). 
\item H. Ujino and M. Wadati, J. Phys. Soc. Jpn. {\bf 65}, 653 (1996); 
      {\it ibid}, 2423 (1996); {\it ibid}, {\bf 66}, 345 (1997); M. Kojima
      and N. Ohta, Nucl. Phys. {\bf B 473}, 455 (1996).
\item M. Vasiliev, Int. J. Mod. Phys. {\bf A 6}, 1115 (1991); 
      A.P. Polychronakos, Phys. Rev. Lett. {\bf 69}, 703 (1992);
      K.H. Hikami and M. Wadati, J. Phys. Soc. Jpn. {\bf 62}, 4203 (1993); 
      H. Ujino and M. Wadati, {\it ibid}, {\bf 63}, 3385 (1994);
      Phys. Rev. Lett. {\bf 73}, 1191 (1994); V. Narayanan and M. Sivakumar, 
      Mod. Phy. Lett {\bf A 11}, 763 (1996).
\item F. Calogero, J. Math. Phys. {\bf 10}, 2191 (1969).
\item T.H. Baker and P.J. Forrester, solv-int/9608004, 9609010; {\it
      ibid}, Nucl. Phys. {\bf B 492}, 682 (1997); J.F. van Dijen,
      q-alg/9609032. 
\item K. Sogo, J. Phys. Soc. Jpn. {\bf 65}, 3097 (1996).
\item I.G. Macdonald, {\it Symmetric functions and Hall polynomials} (Clarendon
      press, Oxford, 1979).
\item A. Cappelli, C.A. Trugenberger and G.R. Zemba, Nucl. Phys. {\bf B 396},
      465 (1993); {\it ibid}, Phys. Rev. {\bf B 306}, 100 (1993); M. Fran, 
      S. Sciuto, A. Lerda and G.R. Zemba, Int. J. Mod. Phys. {\bf A 12}, 4611
      (1997).
\item M.V.N. Murthy, R.K. Bhaduri and D. Sen, Phys. Rev. Lett. {\bf 76},
       4103 (1996).
\item  G.S. Agarwal and S. Chaturvedi, J. Phys. {\bf A 28}, 5747 (1995).
\item  A.P. Polychronakos, Phys. Rev. Lett. {\bf 74}, 5153 (1995).
\item V.G. Kac and A.K. Raina, {\it Bombay Lectures On Highest Weight 
      Representations of Infinite Dimensional Lie Algebras} (Advanced Series
      in Mathematical Physics, Vol. {\bf 2}, World Scientific, Singapore, 
      1987) and references therein.
\end{references}
\end{document}